\documentclass[runningheads]{llncs}
\usepackage{graphicx}
\usepackage{cite}
\usepackage{lineno,hyperref}
\usepackage{setspace}
\usepackage{algpseudocode}
\usepackage{algorithm}
\usepackage{float}
\usepackage{amsmath}
\usepackage[export]{adjustbox}
\usepackage{subfig}

\usepackage{amsthm}
\usepackage{amssymb}
\usepackage{varwidth,comment}
\usepackage{graphicx}
\usepackage{multirow}
\usepackage{pgfplots}
\usetikzlibrary{pgfplots.groupplots}
\usepackage{tikz,array,calc}
\usetikzlibrary{decorations.pathreplacing}
\usepackage{caption}
\begin{document}
\title{A one-phase tree-based algorithm for mining high-utility itemsets from a transaction database
	}
%
%

\author{Siddharth Dawar, Vikram Goyal, Debajyoti Bera}
\institute{Indraprastha Institute of Information Technology, Delhi, India
\email{\{siddharthd,vikram,dbera\}@iiitd.ac.in}}

%

%
\maketitle              
\begin{abstract} 
High-utility itemset mining finds itemsets from a transaction database with utility no less than a fixed user-defined threshold. The utility of an itemset is defined as the sum of the utilities of its item. Several algorithms were proposed to mine high-utility itemsets. However, no state-of-the-art algorithm performs consistently good across dense and sparse datasets. In this paper, we propose a novel data structure called Utility-Tree, and a tree-based algorithm called UT-Miner that mines high-utility itemsets in one-phase only without generating any candidates and uses a lightweight construction method to reduce the cost of creating projected databases during the search space exploration. The transaction information is stored compactly with every node of the Utility-Tree, and the information is computed efficiently during the recursive invocation of the algorithm. Experimental results on several real-life dense and sparse datasets reveal that UT-Miner is among the top-performing efficient algorithms across different datasets. 

\end{abstract}

\section{Introduction}
\label{intro} 
High-utility itemset mining (HUIM) finds itemsets from a transaction database with utility no less than a user-defined threshold. High-utility itemset mining has been used to find the set of profitable products by retail stores for applications like inventory management, shelf-space management, etc. HUIM was applied to find the set of differential expressed genes from gene expression data \cite{biomedical_appln} across different experimental conditions. Kiran et al. \cite{ssdbm} coined the notion of spatial high-utility itemset mining and proposed novel application of HUIM by identifying highly polluted geographical regions for pollution monitoring and monitor congestion at various road segments. The problem of high-utility pattern mining has been studied for different databases like sequences \cite{sequences}, data streams \cite{data_stream}, episodes \cite{epsiodes} and graphs \cite{graphs}.

Several algorithms have been proposed in the literature to mine high-utility itemsets from a transaction database. High-utility itemset mining algorithms represent the transaction database through a summarized data structure and mine high-utility itemsets by recursively constructing projected databases from the global data structure. A projected database represents the transactions that contain a particular prefix itemset to explore. The bottleneck of HUIM algorithms is the exponential search space for exploration and the time spent to construct the projected database during recursive calls. The algorithms can be broadly divided into three classes: tree-based, list-based, and projection-based algorithms based on the data structure used to represent the transaction database. Tree-based algorithms like UP-Growth+ \cite{up_tree}, UP-Hist \cite{uphist} mine itemsets from the database in two phases. In the first phase, the transaction database is stored as a tree structure, and candidate high-utility itemsets are generated by using upper-bound estimates like TWU \cite{up_tree}. In the next phase, another database scan is performed to find high-utility itemsets by computing the utility of candidates. The second phase, called the verification phase, dominates the performance of tree-based algorithms as a large number of candidates get generated for lower thresholds. 

List-based algorithms like HUI-Miner \cite{hui_miner}, FHM \cite{fhm} maintain an inverted-list data structure to mine high-utility itemsets without generating any candidates in one-phase only. List-based algorithms are known to perform better compared to tree-based algorithms. However, the operation to construct the inverted-list for a $\lbrace k \rbrace-$itemset by joining the lists of $\lbrace k-1 \rbrace-$itemsets is a costly operation. List-based algorithms might explore itemsets that are non-existent in the database. Projection-based algorithms like EFIM \cite{efim}, D2HUP \cite{d2hup} etc. were proposed that represent the transaction database as transactions only and utilize several techniques like closure, transaction merging, etc. to mine patterns efficiently in one-phase only. 

Transaction databases can be categorized into dense and sparse datasets based on the number of items and average transaction length. Dense datasets have fewer items and longer transactions compared to sparse datasets. Dense datasets are generated from games like Chess, and species of mushroom that have very few items and longer transaction length. Sparse datasets are generated by retail giants like Walmart, Amazon, etc. that sell millions or billions of products, but a customer usually purchase a few products only.  

EFIM is known to be the most efficient algorithm across dense datasets, and D2HUP is the most efficient algorithm in terms of execution time across sparse datasets. EFIM doesn't perform well across the benchmark sparse datasets as the transaction merging optimization does not work well. Sparse datasets have a large number of transactions compared to dense datasets, and items appear in very few transactions. EFIM performs a binary search to find out the transactions that contain an item during the projected database creation. The effectiveness of transaction merging reduces for sparse datasets and EFIM spends a lot of time doing binary searches during the mining process. D2HUP utilizes the hyperlink structure across transactions to avoid binary search during the creation of the projected database but does not perform well for dense datasets. We need to know whether a dataset is dense or sparse to choose an algorithm that will perform the best for a dataset. The motivation behind designing our proposed data structure and algorithm is to come up with a data structure and an algorithm that performs consistently well across dense and sparse datasets.

We design a data structure that stores the complete information with every node in the tree to compute the utility of an itemset and decide by computing an upper-bound score whether to expand the current itemset or not to search for high-utility itemsets. We store a data structure similar to the utility-list \cite{hui_miner} structure with every node of the tree and propose an algorithm called UT-Miner that extracts high-utility itemsets in one-phase only and generate only valid itemsets present in the database, unlike vertical mining algorithms like HUI-Miner \cite{hui_miner} and FHM \cite{fhm}. Our proposed algorithm creates the projected database by performing minimal changes on the global tree instead of creating the complete local tree to reduce the time for creating projected databases and design a more memory-efficient algorithm. It can be quickly verified from Table \ref{Tab:sp} and Table \ref{Tab:ds} that our proposed algorithm UT-Miner is among the top algorithms ranked in ascending order of the total execution time across dense and sparse datasets as validated from our experiment results in Section \ref{sec:Experiments and Results}.
\begin{table}
	
	\begin{center}
		\caption{Top-3 efficient algorithms on sparse datasets \label{Tab:sp}}
		\begin{tabular}{ l l  l l }
			\hline
			\bfseries{Algorithms/Dataset} & \bfseries{Retail} & \bfseries{Kosarak}& \bfseries{ChainStore}  \\ \hline
			FHM \cite{fhm} &  &  &   \\
			mHUIMiner \cite{mhuiminer} &  &  &  \\
			UFH \cite{hybrid} & $\checkmark$  & $\checkmark$ & $\checkmark$    \\
			UT-Miner & $\checkmark$ & $\checkmark$ & $\checkmark$ \\
			HMINER \cite{hminer} &  & $\checkmark$ & \\
			EFIM \cite{efim} &  &  &   \\
			D2HUP \cite{d2hup} & $\checkmark$ &  & $\checkmark$  \\ \hline
		\end{tabular}

	\end{center}
\end{table}

\begin{table}
	
	\begin{center}
		\caption{Top-3 efficient algorithms on dense datasets \label{Tab:ds}}
		\begin{tabular}{ l l  l l l}
			\hline
			\bfseries{Algorithms/Dataset} & \bfseries{Chess} & \bfseries{Mushroom}& \bfseries{Connect} & \bfseries{Accidents}  \\ \hline
			FHM \cite{fhm} &  &  & &  \\
			mHUIMiner \cite{mhuiminer} &  &  & &  \\
			UFH \cite{hybrid} &  &  &   &  \\
			UT-Miner & $\checkmark$ & $\checkmark$ & $\checkmark$ & $\checkmark$\\
			HMINER \cite{hminer} & $\checkmark$ & $\checkmark$ & $\checkmark$ & $\checkmark$ \\
			EFIM \cite{efim} & $\checkmark$ & $\checkmark$  & $\checkmark$  & $\checkmark$ \\
			D2HUP \cite{d2hup} &  &  &  & \\ \hline
		\end{tabular}

	\end{center}
\end{table}

Our contributions are summarized below.  
\begin{enumerate}
    \item We propose a novel data structure called Utility-Tree that stores the transaction information compactly with each node of the tree to mine high-utility itemsets in one-phase only.
    \item We propose a novel tree-based algorithm called UT-Miner that mines high-utility itemsets without generating any candidates and uses a lightweight method to construct the projected database during the search space exploration.
    \item We conduct extensive experiments on several real dense and sparse datasets to compare the performance of UT-Miner with other algorithms. Our experimental study confirms that UT-Miner is among the top-performing algorithms consistently across sparse and dense datasets compared to other algorithms whose performance depends on the nature of the datasets.
\end{enumerate}
This paper is organized as follows. Section \ref{sec:Related Work} reviews the related work and the problem statement is defined in Section \ref{sec:Background}. We describe our proposed data structure and algorithm in Section \ref{sec:our}. Our extensive experimental study across several dense and sparse datasets is presented in Section \ref{sec:Experiments and Results}, and Section \ref{sec:Conclusion and Future Work} concludes the paper.

\section{Related work}\label{sec:Related Work}
Liu et al. \cite{two_phase} proposed a recursive two-phase breadth-first search algorithm to mine high-utility itemsets from a transaction database. This paper defined an anti-monotonic upper bound called transaction-weighted utility (TWU) to prune the search space as the utility measure is neither monotonic nor anti-monotonic. The two-phase algorithm explores the search space in a level-wise manner. The two-phase algorithm scans the database $k$ times, where $k$ is the length of the longest transaction in the database. Level-wise algorithms are memory intensive as they store the candidate high-utility itemsets at level $k-1$ to compute the candidates for the next level.

Ahmed et al. \cite{ihup_tree} proposed the first tree data structure called IHUP-tree and a recursive depth-first search algorithm called IHUP-Miner to find high-utility itemsets with two database scans only. In the first scan, each transaction is inserted to construct an IHUP-tree. Each node of the IHUP-tree stores the item name, frequency, TWU, and pointers to its parent and child nodes. Potential high-utility itemsets are generated by recursively generating local trees from the global IHUP-tree. Another database scan is performed to compute the utility of candidate itemsets and identify the high-utility itemsets. Tseng et al. \cite{upgrowth} proposed another tree data structure called UP-tree and an algorithm called UP-Growth. UP-tree stores an upper bound called node utility with every node, and UP-Growth employs strategies called DGU, DGN, DLU, and DLN to compute better utility estimates compared to TWU during the candidate generation phase. Tseng et al. \cite{up_tree} proposed another algorithm called UP-Growth+ to reduce the overestimated utilities more effectively. Dawar et al. \cite{uphist} proposed another data structure called UP-Hist tree that stores a histogram of item-quantity with each node to compute better utility estimates compared to UP-tree structure. The tree-based algorithms perform better than level-wise algorithms as they require only two scans of the database and require less memory as they explore the search space in a depth-first manner. However, tree-based algorithms spend the majority of their time during the verification phase as a large number of candidates are generated at lower thresholds.

Liu et al. \cite{hui_miner} proposed a data structure called utility-list and an algorithm called HUI-Miner to mine high-utility itemsets in one-phase only without generating any candidates. The utility-list structure stores tuples that contain a transaction identifier, exact-utility, and remaining-utility for each itemset. Initially, the database is scanned to compute the TWU of items. Items with TWU less than the minimum utility threshold called unpromising items are removed from the database, and utility-list for itemsets containing one item is constructed. The utility-list for 2-itemset (i.e., an itemset containing two items) is constructed by joining the utility-list of its items. The utility-list for a $\lbrace k \rbrace $-itemset with k$>$2 is constructed from the utility-list of two $\lbrace k-1 \rbrace $-itemsets and utility-list of the prefix itemset. Fournier-Viger et al. \cite{fhm} proposed a strategy called EUCP that stores the TWU for every pair of items to reduce the number of join operations performed by HUI-Miner and another list-based algorithm called FHM. Duong et al. \cite{bufferedlist} proposed an algorithm called ULB-Miner that uses a memory buffer to store and retrieve utility-lists efficiently. Peng et al. \cite{mhuiminer} proposed an algorithm called mHUIMiner (modified HUI-Miner) that utilizes a tree structure to avoid considering itemsets that are non-existent in the database. The algorithm integrates the IHUP-tree structure into the original HUI-Miner algorithm. Dawar et al. \cite{hybrid} proposed an algorithm called UFH that combines a tree-based algorithm called UP-Growth+ and a list-based algorithm called FHM to mine high-utility itemsets. Several optimizations like memoization, early termination, and transaction merging were also used to enhance the performance of the UFH algorithm. 

Zida et al. \cite{efim} proposed an algorithm called EFIM that utilizes techniques like database projection and transaction merging using linear time and space implementation. EFIM is, in general, two to three orders of magnitude faster than other algorithms on dense datasets. Liu et al. \cite{d2hup} proposed a linear data structure called CAUL and an algorithm called D2HUP to mine high-utility itemsets. D2HUP employs techniques like closure and lookahead pruning to prune the search space effectively. It performs one to three orders of magnitude better than other algorithms on sparse datasets. Krishnamoorthy \cite{hminer} proposed an algorithm called HMINER that utilizes a compact utility-list data structure and several techniques like transaction merging, and lookahead pruning. The HMINER algorithm constructs the compact utility-list for the promising extensions of the current prefix simultaneously in linear time. Jaysawal et al. \cite{dmhups} proposed a data structure called IUData List similar to the compact utility-list structure proposed by Krishnamoorthy \cite{hminer} and an algorithm with pruning strategies like transaction merging, and lookahead pruning.  

In this paper, we propose a data structure that maps a data structure adapted from the utility-list \cite{hui_miner} data structure on the tree structure \cite{up_tree} and the first one-phase tree-based algorithm to mine high-utility itemsets. The closely related previous works \cite{hybrid,mhuiminer} either switch from a tree structure to a utility-list structure or utilize the tree structure to guide the itemset mining process by creating utility-lists during the mining process. We propose a novel approach where minimal changes are made on the global tree to create a projected database during the mining process, and our proposed approach, unlike the state-of-the-art algorithms, performs consistently well on both dense and sparse datasets as validated from our experimental study.

\section{Preliminaries and problem statement }\label{sec:Background}
Consider a set of items $I=\lbrace i_1,i_2,...,i_m \rbrace$ and a transaction database $D=\lbrace T_1,T_2,...,T_n \rbrace$ where every transaction is a subset of $I$. Every item in a transaction is associated with a positive weight. An example of such a database is shown in Table \ref{Tab:example}. In our example, $I=\lbrace A,B,C,D,E,F,G \rbrace$. The utility of an item $i$ in a transaction $T$ denoted by $u(i,T)$ is the weight associated with the item in $T$. For example, the utility of item $\lbrace F \rbrace$ in $T_1$ is 13. The utility of an itemset $X$ in a transaction $T$ denoted by $u(X,T)$ is the sum of utility of its items. For example, the utility of itemset $\lbrace CF \rbrace$ in $T_1$ is 14. The utility of itemset $X$ in the database is defined as: $u(X) = \sum_{\substack{X \subseteq T \\ T \in D}} u(X,T)$. For example, the utility of the itemset $\lbrace CF \rbrace$ for our example database is 29.\\

\noindent{\bf Problem Statement:}
	An itemset $X$ is called a high-utility itemset if $u(X)$ is no less than a given minimum user-defined threshold denoted by $\theta$. Given a transaction database $D$, and a minimum user-defined threshold $\theta$, the aim is to find all high-utility itemsets.

The utility of an itemset in the database is neither monotone nor anti-monotone, i.e., the superset of a low-utility itemset can be high-utility, and the subset of a high-utility itemset can be low-utility. For example, the utility of $\lbrace C \rbrace$ and $\lbrace CF \rbrace$ for our example database is 14 and 29, respectively. Let $\theta$ be 17. The itemset $\lbrace C \rbrace$ is not a high-utility itemset, but its superset $\lbrace CF \rbrace$ is a high-utility itemset. The utility of $\lbrace A \rbrace$ and $\lbrace ACD \rbrace$ is 5 and 19, respectively. Even though $\lbrace ACD \rbrace$ is a high-utility itemset, the subset $\lbrace A \rbrace$ is not a high-utility itemset. The search space for the high-utility itemset mining problem is exponential in the number of items. 

Liu et al. \cite{two_phase} defined an upper-bound called transaction-weighted utility (TWU) \cite{twdc} that satisfies the anti-monotonicity property. The transaction utility (TU) for a transaction is defined as the sum of utility if its items. For example, the transaction utility of $T_1$ in Table \ref{Tab:example} is 18. The transaction-weighted utility of an itemset $X$ is the sum of $TU$ of transactions that contain $X$. For example, the TWU of $\lbrace A \rbrace$ is 37. Let $\theta$ be 40. Itemset $\lbrace A \rbrace$ and its supersets can't be high-utility itemsets as its TWU is less than $\theta$. Liu et al. \cite{hui_miner} proposed another anti-monotonic upper bound called EU-RU that is tighter than TWU to prune the search space. The exact-utility (EU) of an itemset in a transaction is the utility of the itemset in the transaction. The remaining-utility (RU) of an itemset X in a transaction is the sum of utility of items that appear after X in the transaction. The items in a transaction are ordered according to a predefined ordering like lexicographic etc. If the sum of exact-utility and remaining-utility (EU-RU) of an itemset X in the transactions containing X is less than $\theta$, X and its supersets can't be high-utility. For example, the exact-utility and remaining-utility of $\lbrace CF \rbrace$ in $T_1$ are 14, and 3, respectively, assuming that the items in every transaction are sorted in lexicographic order.    

\begin{table}
	
	\begin{center}
		\caption{$Example \, database$ \label{Tab:example}}
		\begin{tabular}{  l l l }
				\hline 
				\bfseries{TID} & \bfseries{Transaction} & \bfseries{TU} \\ \hline
				$T_1$ & $(C:1) \, (E:1) \, (F:13) \, (G:3)$ & 18\\ 
				$T_2$ & $(B:1) \, (D:1) \, (F:6) \, (G:6)$ & 14  \\ $T_3$ & $(B:2) \, (C:4) \, (F:4) \, (G:3)$ & 13 \\  
				$T_4$ & $(B:1) \, (D:1) \, (E:1) \, (F:1) \, (G:1)$ & 5\\ 
				$T_5$ & $(B:1) \, (C:1) \, (E:1) \, (F:1) \, (G:1)$ & 5\\ 
				$T_6$ & $(B:10) \, (E:1) \, (F:1) \, (G:1)$ & 13  \\ $T_7$ & $(A:5) \, (C:4) \, (D:10) \, (E:12) \, (G:6)$ & 37  \\ 
				$T_8$ & $(D:5) \, (E:2) \, (F:10)$ & 17  \\ 
				$T_9$ & $(C:4) \, (D:5) \, (E:2) \, (F:1)$ & 12  \\ 
				$T_{10}$ & $(F:15) \, (G:10)$ & 25  \\ 
				
				\hline    
			\end{tabular}
			
	\end{center}
\end{table}

\section{Our proposed data structure and tree-based algorithm}\label{sec:our} 
In this section, we propose a new data structure called Utility-tree and a one-phase tree-based algorithm called UT-Miner. UT-Miner creates a lightweight projected database on the global Utility-tree to reduce the cost of creating projected databases during recursive calls.

\subsection{Utility-tree data structure}
Each node N of the Utility tree stores the following information: 1) item name N.$item$, 2) a HashMap of key-value pairs N.$gmap$ with every entry of the form $\langle$ tid, (exact-utility, remaining-utility) $\rangle$, 3) a linked-list of local nodes N.$local\_list$, 4) a unique identifier N.$id$, 5) a pointer to the parent node N.$parent$, 6) a pointer N.$hlink$ to the node which has the same name as N.$item$. A local node L stores the following information: 1) prefix id, 2) a HashMap of key-value pairs L.$lmap$ with every entry of the form $\langle$ id, (extension-utility, remaining-utility, prefix-utility) $\rangle$. The prefix-utility is the utility of the prefix for the identifier (id). The exact-utility of an itemset for an identifier (id) is the sum of extension-utility and prefix-utility. The root of the Utility-tree is a special node that points to its child nodes. A header table is maintained with the Utility-tree for efficient traversal. The header table stores the following information: 1) item name, 2) TWU, 3) a pointer $link$. The nodes along a path in the Utility-tree are maintained in descending order of their TWU values. All nodes with the same label are stored in a linked-list, and $link$ pointer points to the head node in the list.        
\begin{figure}
	\begin{center}
	
		\includegraphics[width=12cm]{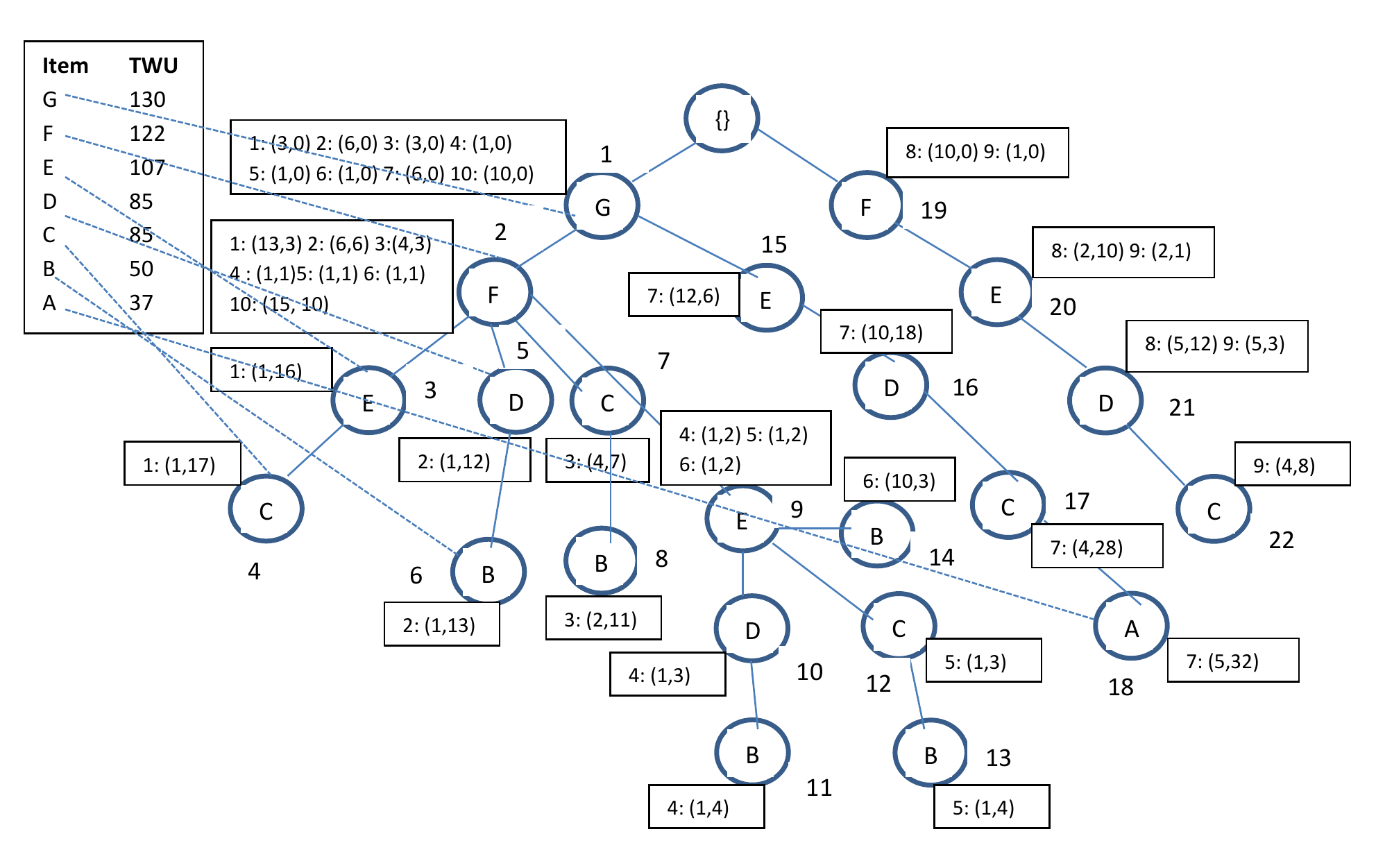}
	\end{center}
	\caption{Global Utility-tree}
	\label{fig:utility_tree}
\end{figure}

Now, we discuss the process to construct the Utility-tree from a transaction database for a user-defined minimum utility threshold $\theta$. Initially, the database is scanned to compute TWU for items. The items with TWU less than $\theta$ called unpromising items are identified as such items can not be a part of any high-utility itemset due to the anti-monotonicity property of the TWU bound. The unpromising items are removed, and transactions are arranged in lexicographic order \cite{efim}. The empty transactions are removed, and transaction merging \cite{efim} is performed. Another database scan is performed to sort the items in a transaction according to increasing order of TWU and compute exact-utility, and remaining-utility for each item. Every transaction is inserted one by one to construct the utility-tree. The header table contains only items with EU-RU no less than $\theta$. The nodes along a path from the root to a leaf node in the Utility-tree and items in the header table are sorted in decreasing order of TWU only. The tree construction process is similar to the one followed by other tree-based algorithms \cite{up_tree,uphist,ihup_tree}. The $local\_list$ associated with each node is initially empty. 

\subsection{UT-Miner algorithm}
\begin{algorithm}[!h]
	\caption{UT-Miner ($\alpha$,$T$,$H$,$hlist$,$\theta$)}
	\textbf{Input:} Prefix $\alpha$ (initially empty), Utility-tree $T$, a header table for $T$, list of extensions to explore, $\theta$: a user-specified threshold. \\
	\textbf{Output:} the set of high-utility itemsets with $\alpha$ as prefix.
	\label{algo}
	\begin{algorithmic}[1]
		\For{each entry $\lbrace i \rbrace$ in $hlist$}
		        \State Itemset $I=\alpha \cup i$. \Comment{Append the extension i to the current prefix.}
		  		\State Compute sumEU and sumRU for $I$  by following the links from the header table $H$ for item $\lbrace i \rbrace$. \Comment{Compute sum of the exact-utility and remaining-utility of $I$ for the transactions containing $I$. }
		  		\If{$I.sumEU \geq \theta$}
		  		\State Output $I$ as a high-utility itemset.
		  		\EndIf
		  		\State Compute TWU for the extensions i.e. ancestors of $I$ in $T$ and identify unpromising items $ulist$.
		  		\State Initialize variable $ub = I.sumEU+I.sumRU$.
		  		\State Remove the contribution of items in $ulist$ from $ub$. Call the updated bound as $updatedub$.
		  		\If{$updatedub < \theta$}
		  		\State Return.
		  		\EndIf
		  		\State Add a node in $local\_list$ for every ancestor of $I$ in $T$.
		  		\State Construct the list of extensions for I denoted by $hlist_I$.
		  		\State Call UT-Miner($I$,$T$,$H$,$hlist_I$,$\theta$).
		  		
		 \EndFor
		 \State Remove the node associated with $\alpha$ from $local\_list$ associated with ancestors of $\alpha$ in $T$. 
	\end{algorithmic}
\end{algorithm}
Our proposed algorithm, called UT-Miner (Algorithm \ref{algo}), takes as input a prefix, a Utility-tree constructed from the transaction database as described above, a header table associated with the Utility-tree, a list of extensions to explore for the current prefix, and the minimum utility threshold. UT-Miner returns the complete set of high-utility itemsets. Initially, UT-Miner is called for an empty prefix with all the items present in the header table as extensions for exploration. The items in the header table are explored in a bottom-up manner. An itemset is created by adding an item from the header table to the current prefix (line 2). The exact-utility and remaining-utility for the itemset are computed by traversing the linked-list associated with the item from the header table (line 3). The itemset is output as high-utility (line 4) if its utility is no less than the minimum utility threshold. The TWU for the ancestors (i.e., items above item $i$ in the tree) of item $i$ is computed to identify the unpromising items (line 7). An item is unpromising if its TWU is less than the minimum utility threshold. Such items can't be a part of any high-utility itemset. The utility of unpromising items is removed from the EU-RU bound for the current itemset $I$ to compute a tighter upper bound score. If the new bound is less than the threshold, no further extensions need to be explored (line 10). Else, a node will be created in the $local\_list$ of all ancestors of $i$ in the Utility-tree (line 13). The strategy to remove the unpromising items and compute a tighter bound is inspired by the DLU \cite{up_tree} strategy, and the proof of correctness can be referred from \cite{hui_miner}. The list of ancestors with EU-RU bound no less than the threshold is inserted in the list of extensions to explore further (line 14), and the UT-Miner algorithm is called recursively. The node in the $local\_list$ is removed from all ancestors after the complete processing for the current prefix $\alpha$ (line 17).      

\begin{figure}
	\begin{center}
	
		\includegraphics[width=12cm]{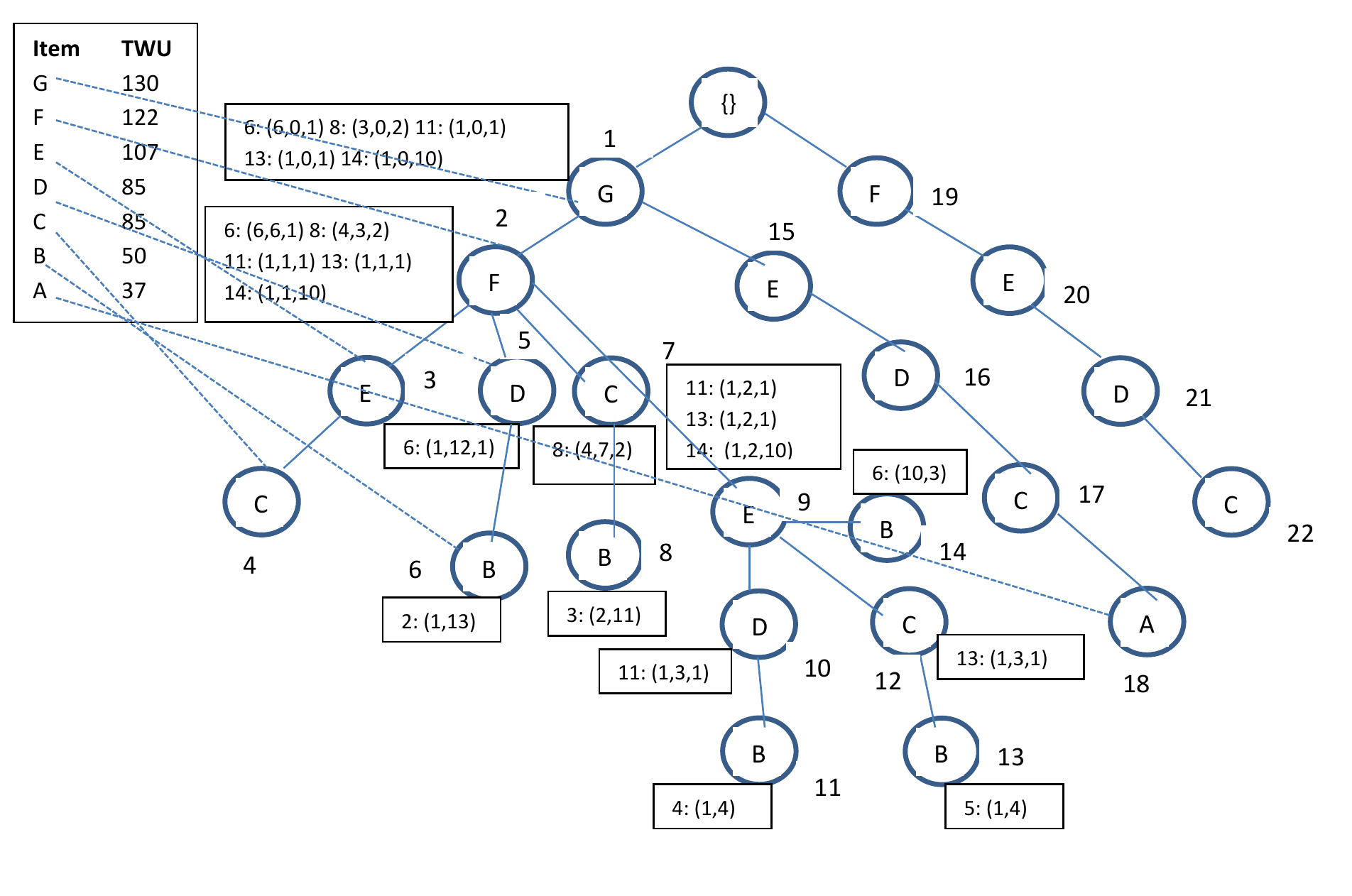}
	\end{center}
	\caption{Global Utility-tree with $local\_list$ for prefix $\lbrace B \rbrace$. The $gmap$ associated with the ancestors of $\lbrace B \rbrace$ and other nodes in the Utility-tree is not shown. }
	\label{fig:utility_tree_2}
\end{figure}

Now, we will illustrate the execution of our algorithm through an example. Consider a transaction database, as shown in Table \ref{Tab:example}, and let the minimum utility threshold be 20. The database is scanned to compute the TWU of items, as shown in Table \ref{Tab:twu}.
\begin{table}[!h]
	
	\begin{center}
		\caption{TWU of items \label{Tab:twu}}
		\begin{tabular}{  l  l  l  l  l  l  l  l  l }
		\hline
		\bfseries{Item} & \bfseries{A} & \bfseries{B} & \bfseries{C} & \bfseries{D} & \bfseries{E} & \bfseries{F} & \bfseries{G} \\ \hline
		\bfseries{TWU} & 37 & 50 & 85 & 85 & 107 & 122 & 130 \\
		\hline
	     \end{tabular}
	\end{center}
\end{table}
There are no unpromising items in this example. The items in every transaction are sorted in increasing order of TWU, and transaction merging is performed. The reorganized transaction is inserted to form the global Utility-tree, as shown in Figure \ref{fig:utility_tree}. Let us observe the processing for prefix $\lbrace B \rbrace$. The first node corresponding to $\lbrace B \rbrace$ is processed from the header table, and the linked-list corresponding to $\lbrace B \rbrace$ is traversed to compute the EU-RU bound for $\lbrace B \rbrace$. The EU-RU bound corresponding to $\lbrace B \rbrace$ is 50. Prefix $\lbrace B \rbrace$ will be processed further as its EU-RU is more than the minimum utility threshold. The UT-Miner algorithm will traverse the nodes corresponding to $\lbrace B \rbrace$ again from the header table, and compute the TWU for items which are between $\lbrace B \rbrace$ and the root along every path from $\lbrace B \rbrace$ to the root node. The TWU of items for prefix $\lbrace B \rbrace$ is shown in Table \ref{Tab:B}.
\begin{table}[!h]
	
	\begin{center}
		\caption{TWU of items for the prefix $\lbrace B \rbrace$ \label{Tab:B}}
		\begin{tabular}{  l  l  l  l  l  l  l }
		\hline
		\bfseries{Item} &  \bfseries{C} & \bfseries{D} & \bfseries{E} & \bfseries{F} & \bfseries{G}  \\ \hline
		\bfseries{TWU} & 18 & 19 & 23 & 50 & 50 \\
		\hline
	 \end{tabular}
	\end{center}
\end{table}
Item $\lbrace C \rbrace$ and $\lbrace D \rbrace$ are unpromising for prefix $\lbrace B \rbrace$. The EU values for $\lbrace C \rbrace$ and $\lbrace D \rbrace$ will be removed from the EU-RU bound for $\lbrace B \rbrace$. The updated bound after removing the effect of unpromising items is 43. Therefore, prefix $\lbrace B \rbrace$ can be explored further. The linked-list associated with prefix $\lbrace B \rbrace$ is again processed from the header table of Utility-tree to construct the $local\_list$ for all the ancestors (i.e., nodes between $\lbrace B \rbrace$ and the root node) for every $\lbrace B \rbrace$ node in the linked-list. For every $\lbrace B \rbrace$ node, its exact-utility is computed, and a pair $\langle ID, PU \rangle$ along with the transaction id's associated with $\lbrace B \rbrace$ is sent to all ancestors. $ID$ is the unique id associated with every node of the Utility-tree. PU is the prefix utility, i.e., the utility of $\lbrace B \rbrace$. The $local\_list$ created for the ancestors of prefix $\lbrace B \rbrace$ is shown in Figure \ref{fig:utility_tree_2}.

The valid extensions for prefix $\lbrace B \rbrace$ are $\lbrace E \rbrace$, $\lbrace F \rbrace$, and $\lbrace G \rbrace$. The EU-RU bound for $\lbrace BE \rbrace$, $\lbrace BF \rbrace$, and $\lbrace BG \rbrace$ is 21, 40, and 27 respectively. Let us focus on the processing for prefix $\lbrace BE \rbrace$. The ancestors (i.e. valid extensions) for $\lbrace BE \rbrace$ are $F$ and $G$ respectively. The TWU of $F$ and $G$ is 21 only for the prefix $\lbrace BE \rbrace$ as there is a single path from $E$ to the root. No item is unpromising, and a node in the $local\_list$ of $F$, and $G$ for the prefix $\lbrace BE \rbrace$ will be created. For every $\lbrace E \rbrace$ node, its exact-utility is computed and a pair $\langle ID, PU \rangle$, along with the transaction id's associated with $\lbrace BE \rbrace$ is sent to all ancestors, i.e., $F$ and $G$. The PU is 15 for the prefix $\lbrace BE \rbrace$. A node for the prefix $\lbrace BE \rbrace$ will be added in the $local\_list$ of $F$, and $G$, respectively. After the complete processing for a prefix, the node inserted in the $local\_list$ of its ancestors has to be removed before processing the next prefix.          

\section{Experiments and results}\label{sec:Experiments and Results}
In this section, we compare the performance of UT-Miner algorithm with FHM \cite{fhm}, mHUIMiner \cite{mhuiminer}, UFH \cite{hybrid}, EFIM \cite{efim}, D2HUP \cite{d2hup}, and HMINER \cite{hminer}. The source code for the algorithms was obtained from SPMF \cite{spmf} library. The total execution time, the number of generated candidates, and the main memory consumed during execution are used as the performance metrics. 

The experiments were performed on an Intel Xeon(R) CPU=26500@2.00 GHz with 16 GB RAM and Windows Server 2012 operating system on sparse and dense datasets obtained from the SPMF library \cite{spmf}. The datasets vary in the number of transactions, the number of items, and the average transaction length, as shown in Table \ref{Tab:datasets}. The internal utility values were generated from a uniform distribution in the range from 1 to 10. The external utility values were generated from a Gaussian distribution. Only the ChainStore dataset contains real utility values.
\begin{table}
	
	\begin{center}
		\caption{$ Characteristics \, of \, real \, datasets$ \label{Tab:datasets}}
		\begin{tabular}{ l l  l l  l}
			\hline
			\bfseries{Dataset} & \bfseries{\#Tx} & \bfseries{Avg. length}& \bfseries{\#Items} &  \bfseries{Type}  \\ \hline
			Retail & 88,162 & 10.3 & 16,470 & Sparse \\
			Kosarak & 9,90,002 & 8.1 & 41,270 & Sparse \\
			Chainstore & 11,12,949 & 7.2 & 46,086 & Sparse  \\
			Chess & 3,196 & 37 & 75  & Dense \\
			Mushroom & 8,124 & 23 & 119 & Dense \\
			Connect & 67,557 & 43 & 129 & Dense \\
			Accidents & 3,40,183 & 33.8 & 468 & Dense \\ \hline
		\end{tabular}

	\end{center}
\end{table}

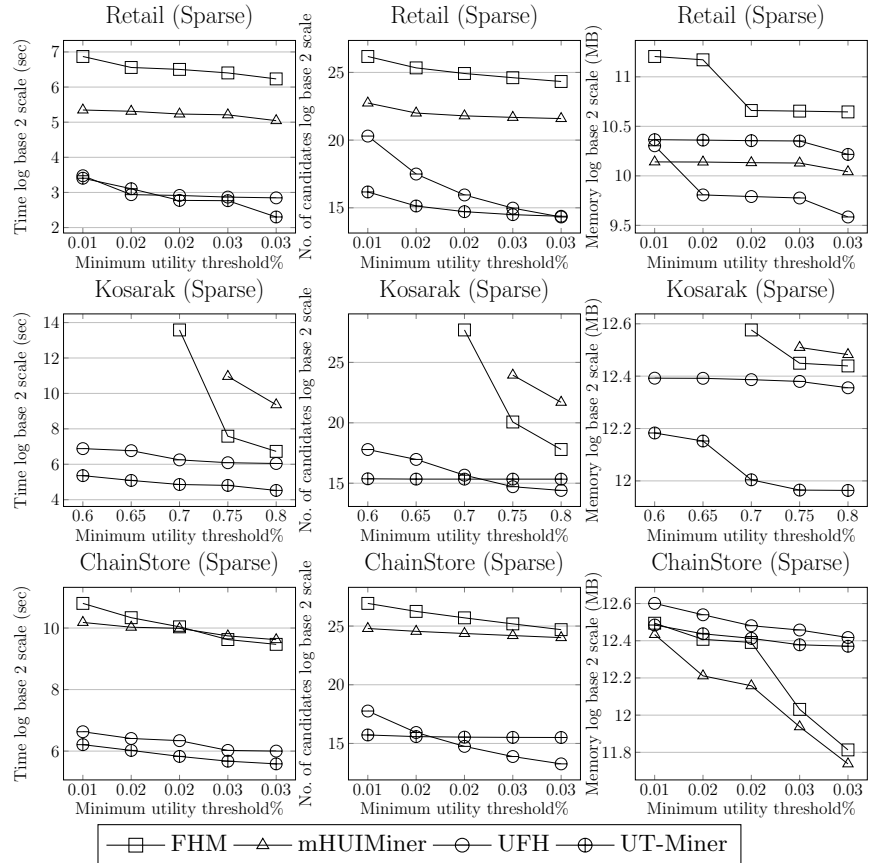
\begin{figure}
	\begin{center}
		\begin{tabular}{ccc}
			\centering
			
			\begin{minipage}{.27\linewidth}
				\begin{tikzpicture}[scale=0.45]
				\begin{axis}[
				title={\huge Retail (Sparse)},
				xlabel={Minimum utility threshold$\%$ },
				ylabel={Time log base 2 scale (sec)},
				enlarge y limits=true,
					legend to name=t1,
					legend image post style={scale=1.3},
				legend style={at={($(0,0)+(1cm,1cm)$)},
					legend columns=4,fill=none,draw=black,anchor=center,align=center},
				ymajorgrids=true,
				mark size=5pt,
				label style={font=\Large},
                tick label style={font=\Large},
                scaled ticks=false, 
                tick label style={/pgf/number format/fixed}
				]
				
				\addplot[
				color=black,
				mark=square,
				style=solid,
				]
				coordinates {
					(0.01,{log2(117)})(0.015,{log2(94.2)})(0.02,{log2(90.7)})(0.025,{log2(84.6)})(0.03,{log2(75.2)})
				};
				\addlegendentry{FHM}
				\addplot[
				color=black,
				mark=triangle,
				style=solid,
				]
				coordinates {
					(0.01,{log2(40.7)})(0.015,{log2(39.7)})(0.02,{log2(37.6)})(0.025,{log2(37)})(0.03,{log2(33)})
				};    
				\addlegendentry{mHUIMiner}
				\addplot[
				color=black,
				mark=halfcircle,
				style=solid,
				]
				coordinates {
					(0.01,{log2(11.12)})(0.015,{log2(7.66)})(0.02,{log2(7.54)})(0.025,{log2(7.3)})(0.03,{log2(7.19)})
				};    
				\addlegendentry{UFH}
				\addplot[
				color=black,
				mark=oplus,
				style=solid,
				]
				coordinates {
					(0.01,{log2(10.59)})(0.015,{log2(8.61)})(0.02,{log2(6.84)})(0.025,{log2(6.8)})(0.03,{log2(4.93)})
				};    
				\addlegendentry{UT-Miner}
				
				\end{axis}
				
				\end{tikzpicture}
			\end{minipage}
			\hspace{0.3cm}
			
			&
			
			\begin{minipage}{.27\linewidth}
				\begin{tikzpicture}[scale=0.45]
				\begin{axis}[
				title={\huge Retail (Sparse)},
				xlabel={Minimum utility threshold$\%$ },
				ylabel={No. of candidates log base 2 scale},
				enlarge y limits=true,
					legend to name=t1,
					legend image post style={scale=1.3},
				legend style={at={($(0,0)+(1cm,1cm)$)},
					legend columns=4,fill=none,draw=black,anchor=center,align=center},
				ymajorgrids=true,
				mark size=5pt,
				label style={font=\Large},
                tick label style={font=\Large},
                scaled ticks=false, 
                tick label style={/pgf/number format/fixed}
				]
				
				\addplot[
				color=black,
				mark=square,
				style=solid,
				]
				coordinates {
					(0.01,{log2(75796543)})(0.015,{log2(42323955)})(0.02,{log2(31756941)})(0.025,{log2(25599800)})(0.03,{log2(21083208)})
				};
				\addlegendentry{FHM}
				\addplot[
				color=black,
				mark=triangle,
				style=solid,
				]
				coordinates {
					(0.01,{log2(6997961)})(0.015,{log2(4192604)})(0.02,{log2(3629588)})(0.025,{log2(3357201)})(0.03,{log2(3158529)})
				};    
				\addlegendentry{mHUIMiner}
				\addplot[
				color=black,
				mark=halfcircle,
				style=solid,
				]
				coordinates {
					(0.01,{log2(1289419)})(0.015,{log2(184640)})(0.02,{log2(63257)})(0.025,{log2(32247)})(0.03,{log2(20497)})
				};    
				\addlegendentry{UFH}
				\addplot[
				color=black,
				mark=oplus,
				style=solid,
				]
				coordinates {
					(0.01,{log2(73788)})(0.015,{log2(35940)})(0.02,{log2(26895)})(0.025,{log2(23098)})(0.03,{log2(21218)})
				};    
				\addlegendentry{UT-Miner}
				
				\end{axis}
				
				\end{tikzpicture}
			\end{minipage}
			\hspace{0.3cm}
			&
			\begin{minipage}{.27\linewidth}
				\begin{tikzpicture}[scale=0.45]
				\begin{axis}[
				title={\huge Retail (Sparse)},
				xlabel={Minimum utility threshold$\%$ },
				ylabel={Memory log base 2 scale (MB)},
				enlarge y limits=true,
					legend to name=t1,
					legend image post style={scale=1.3},
				legend style={at={($(0,0)+(1cm,1cm)$)},
					legend columns=4,fill=none,draw=black,anchor=center,align=center},
				ymajorgrids=true,
				mark size=5pt,
				label style={font=\Large},
                tick label style={font=\Large},
                scaled ticks=false, 
                tick label style={/pgf/number format/fixed}
				]
				
				\addplot[
				color=black,
				mark=square,
				style=solid,
				]
				coordinates {
					(0.01,{log2(2361)})(0.015,{log2(2307)})(0.02,{log2(1618)})(0.025,{log2(1611)})(0.03,{log2(1602)})
				};
				\addlegendentry{FHM}
				\addplot[
				color=black,
				mark=triangle,
				style=solid,
				]
				coordinates {
					(0.01,{log2(1129)})(0.015,{log2(1128)})(0.02,{log2(1123)})(0.025,{log2(1120)})(0.03,{log2(1054)})
				};    
				\addlegendentry{mHUIMiner}
				\addplot[
				color=black,
				mark=halfcircle,
				style=solid,
				]
				coordinates {
					(0.01,{log2(1264)})(0.015,{log2(896)})(0.02,{log2(886)})(0.025,{log2(877)})(0.03,{log2(768)})
				};    
				\addlegendentry{UFH}
				\addplot[
				color=black,
				mark=oplus,
				style=solid,
				]
				coordinates {
					(0.01,{log2(1319)})(0.015,{log2(1315)})(0.02,{log2(1310)})(0.025,{log2(1307)})(0.03,{log2(1190)})
				};    
				\addlegendentry{UT-Miner}
				
				\end{axis}
				
				\end{tikzpicture}
			\end{minipage}
			\\
			\begin{minipage}{.27\linewidth}
				\begin{tikzpicture}[scale=0.45]
				\begin{axis}[
				title={\huge Kosarak (Sparse)},
				xlabel={Minimum utility threshold$\%$ },
				ylabel={Time log base 2 scale (sec)},
				enlarge y limits=true,
					legend to name=t1,
					legend image post style={scale=1.3},
				legend style={at={($(0,0)+(1cm,1cm)$)},
					legend columns=4,fill=none,draw=black,anchor=center,align=center},
				ymajorgrids=true,
				mark size=5pt,
				label style={font=\Large},
                tick label style={font=\Large} 
				]
				
				\addplot[
				color=black,
				mark=square,
				style=solid,
				]
				coordinates {
					(0.7,{log2(12275)})(0.75,{log2(192)})(0.8,{log2(106)})
				};
				\addlegendentry{FHM}
				\addplot[
				color=black,
				mark=triangle,
				style=solid,
				]
				coordinates {
					(0.75,{log2(1992)})(0.8,{log2(657)})
				};    
				\addlegendentry{mHUIMiner}
				\addplot[
				color=black,
				mark=halfcircle,
				style=solid,
				]
				coordinates {
					(0.6,{log2(118)})(0.65,{log2(109)})(0.7,{log2(76)})(0.75,{log2(68)})(0.8,{log2(66)})
				};    
				\addlegendentry{UFH}
				\addplot[
				color=black,
				mark=oplus,
				style=solid,
				]
				coordinates {
					(0.6,{log2(41)})(0.65,{log2(34)})(0.7,{log2(29)})(0.75,{log2(28)})(0.8,{log2(23)})
				};    
				\addlegendentry{UT-Miner}
				
				\end{axis}
				
				\end{tikzpicture}
			\end{minipage}
			\hspace{0.3cm}
			&
			\begin{minipage}{.27\linewidth}
				\begin{tikzpicture}[scale=0.45]
				\begin{axis}[
				title={\huge Kosarak (Sparse)},
				xlabel={Minimum utility threshold$\%$ },
				ylabel={No. of candidates log base 2 scale},
				enlarge y limits=true,
					legend to name=t1,
					legend image post style={scale=1.3},
				legend style={at={($(0,0)+(1cm,1cm)$)},
					legend columns=4,fill=none,draw=black,anchor=center,align=center},
				ymajorgrids=true,
				mark size=5pt,
				label style={font=\Large},
                tick label style={font=\Large} 
				]
				
				\addplot[
				color=black,
				mark=square,
				style=solid,
				]
				coordinates {
					(0.7,{log2(216040928)})(0.75,{log2(1099153)})(0.8,{log2(226796)})
				};
				\addlegendentry{FHM}
				\addplot[
				color=black,
				mark=triangle,
				style=solid,
				]
				coordinates {
					(0.75,{log2(16174405)})(0.8,{log2(3403143)})
				};    
				\addlegendentry{mHUIMiner}
				\addplot[
				color=black,
				mark=halfcircle,
				style=solid,
				]
				coordinates {
					(0.6,{log2(226327)})(0.65,{log2(128397)})(0.7,{log2(51979)})(0.75,{log2(26820)})(0.8,{log2(21681)})
				};    
				\addlegendentry{UFH}
				\addplot[
				color=black,
				mark=oplus,
				style=solid,
				]
				coordinates {
					(0.6,{log2(42208)})(0.65,{log2(41495)})(0.7,{log2(41463)})(0.75,{log2(41434)})(0.8,{log2(41409)})
				};    
				\addlegendentry{UT-Miner}
				
				\end{axis}
				
				\end{tikzpicture}
			\end{minipage}
			\hspace{0.3cm}
			&
			\begin{minipage}{.27\linewidth}
				\begin{tikzpicture}[scale=0.45]
				\begin{axis}[
				title={\huge Kosarak (Sparse)},
				xlabel={Minimum utility threshold$\%$ },
				ylabel={Memory log base 2 scale (MB)},
				enlarge y limits=true,
					legend to name=t1,
					legend image post style={scale=1.3},
				legend style={at={($(0,0)+(1cm,1cm)$)},
					legend columns=4,fill=none,draw=black,anchor=center,align=center},
				ymajorgrids=true,
				mark size=5pt,
				label style={font=\Large},
                tick label style={font=\Large} 
				]
				
				\addplot[
				color=black,
				mark=square,
				style=solid,
				]
				coordinates {
					(0.7,{log2(6108)})(0.75,{log2(5593)})(0.8,{log2(5553)})
				};
				\addlegendentry{FHM}
				\addplot[
				color=black,
				mark=triangle,
				style=solid,
				]
				coordinates {
					(0.75,{log2(5832)})(0.8,{log2(5723)})
				};    
				\addlegendentry{mHUIMiner}
				\addplot[
				color=black,
				mark=halfcircle,
				style=solid,
				]
				coordinates {
					(0.6,{log2(5376)})(0.65,{log2(5374)})(0.7,{log2(5355)})(0.75,{log2(5330)})(0.8,{log2(5241)})
				};    
				\addlegendentry{UFH}
				\addplot[
				color=black,
				mark=oplus,
				style=solid,
				]
				coordinates {
					(0.6,{log2(4650)})(0.65,{log2(4552)})(0.7,{log2(4108)})(0.75,{log2(3998)})(0.8,{log2(3995)})
				};    
				\addlegendentry{UT-Miner}
				
				\end{axis}
				
				\end{tikzpicture}
			\end{minipage}
			\\ 
			\begin{minipage}{.27\linewidth}
				\begin{tikzpicture}[scale=0.45]
				\begin{axis}[
				title={\huge ChainStore (Sparse)},
				xlabel={Minimum utility threshold$\%$ },
				ylabel={Time log base 2 scale (sec)},
				enlarge y limits=true,
					legend to name=t1,
					legend image post style={scale=1.3},
				legend style={at={($(0,0)+(1cm,1cm)$)},
					legend columns=4,fill=none,draw=black,anchor=center,align=center},
				ymajorgrids=true,
				mark size=5pt,
				label style={font=\Large},
                tick label style={font=\Large},
                scaled ticks=false, 
                tick label style={/pgf/number format/fixed}
				]
				
				\addplot[
				color=black,
				mark=square,
				style=solid,
				]
				coordinates {
					(0.01,{log2(1784)})(0.015,{log2(1293)})(0.02,{log2(1052)})(0.025,{log2(790)})(0.03,{log2(709)})
				};
				\addlegendentry{FHM}
				\addplot[
				color=black,
				mark=triangle,
				style=solid,
				]
				coordinates {
					(0.01,{log2(1162)})(0.015,{log2(1043)})(0.02,{log2(1017)})(0.025,{log2(858)})(0.03,{log2(788)})
				};    
				\addlegendentry{mHUIMiner}
				\addplot[
				color=black,
				mark=halfcircle,
				style=solid,
				]
				coordinates {
					(0.01,{log2(99)})(0.015,{log2(85)})(0.02,{log2(81)})(0.025,{log2(65)})(0.03,{log2(64)})
				};    
				\addlegendentry{UFH}
				\addplot[
				color=black,
				mark=oplus,
				style=solid,
				]
				coordinates {
					(0.01,{log2(74)})(0.015,{log2(65)})(0.02,{log2(56.8)})(0.025,{log2(51)})(0.03,{log2(48)})
				};    
				\addlegendentry{UT-Miner}
				
				\end{axis}
				
				\end{tikzpicture}
			\end{minipage}
			\hspace{0.3cm}
			&
			\begin{minipage}{.27\linewidth}
				\begin{tikzpicture}[scale=0.45]
				\begin{axis}[
				title={\huge ChainStore (Sparse)},
				xlabel={Minimum utility threshold$\%$ },
				ylabel={No. of candidates log base 2 scale},
				enlarge y limits=true,
					legend to name=t1,
					legend image post style={scale=1.3},
				legend style={at={($(0,0)+(1cm,1cm)$)},
					legend columns=4,fill=none,draw=black,anchor=center,align=center},
				ymajorgrids=true,
				mark size=5pt,
				label style={font=\Large},
                tick label style={font=\Large},
                scaled ticks=false, 
                tick label style={/pgf/number format/fixed}
				]
				
				\addplot[
				color=black,
				mark=square,
				style=solid,
				]
				coordinates {
					(0.01,{log2(127966995)})(0.015,{log2(79766573)})(0.02,{log2(54505512)})(0.025,{log2(38188741)})(0.03,{log2(27156627)})
				};
				\addlegendentry{FHM}
				\addplot[
				color=black,
				mark=triangle,
				style=solid,
				]
				coordinates {
					(0.01,{log2(29055101)})(0.015,{log2(24559068)})(0.02,{log2(21621033)})(0.025,{log2(19147978)})(0.03,{log2(16952254)})
				};    
				\addlegendentry{mHUIMiner}
				\addplot[
				color=black,
				mark=halfcircle,
				style=solid,
				]
				coordinates {
					(0.01,{log2(222065)})(0.015,{log2(62553)})(0.02,{log2(27581)})(0.025,{log2(15141)})(0.03,{log2(9810)})
				};    
				\addlegendentry{UFH}
				\addplot[
				color=black,
				mark=oplus,
				style=solid,
				]
				coordinates {
					(0.01,{log2(54056)})(0.015,{log2(49078)})(0.02,{log2(47541)})(0.025,{log2(46890)})(0.03,{log2(46570)})
				};    
				\addlegendentry{UT-Miner}
				
				\end{axis}
				
				\end{tikzpicture}
			\end{minipage}
			\hspace{0.3cm}
			&
			\begin{minipage}{.27\linewidth}
				\begin{tikzpicture}[scale=0.45]
				\begin{axis}[
				title={\huge ChainStore (Sparse)},
				xlabel={Minimum utility threshold$\%$ },
				ylabel={Memory log base 2 scale (MB)},
				enlarge y limits=true,
					legend to name=t1,
					legend image post style={scale=1.3},
				legend style={at={($(0,0)+(1cm,1cm)$)},
					legend columns=4,fill=none,draw=black,anchor=center,align=center},
				ymajorgrids=true,
				mark size=5pt,
				label style={font=\Large},
                tick label style={font=\Large},
                scaled ticks=false, 
                tick label style={/pgf/number format/fixed}
				]
				
				\addplot[
				color=black,
				mark=square,
				style=solid,
				]
				coordinates {
					(0.01,{log2(5770)})(0.015,{log2(5433)})(0.02,{log2(5371)})(0.025,{log2(4185)})(0.03,{log2(3599)})
				};
				\addlegendentry{FHM}
				\addplot[
				color=black,
				mark=triangle,
				style=solid,
				]
				coordinates {
					(0.01,{log2(5532)})(0.015,{log2(4742)})(0.02,{log2(4570)})(0.025,{log2(3923)})(0.03,{log2(3416)})
				};    
				\addlegendentry{mHUIMiner}
				\addplot[
				color=black,
				mark=halfcircle,
				style=solid,
				]
				coordinates {
					(0.01,{log2(6212)})(0.015,{log2(5954)})(0.02,{log2(5716)})(0.025,{log2(5625)})(0.03,{log2(5470)})
				};    
				\addlegendentry{UFH}
				\addplot[
				color=black,
				mark=oplus,
				style=solid,
				]
				coordinates {
					(0.01,{log2(5731)})(0.015,{log2(5547)})(0.02,{log2(5453)})(0.025,{log2(5324)})(0.03,{log2(5295)})
				};    
				\addlegendentry{UT-Miner}
				
				\end{axis}
				
				\end{tikzpicture}
			\end{minipage}
			\\
			
		\end{tabular}
		
		\ref{t1}
		\caption{Performance evaluation for FHM, mHUIMiner, UFH and UT-Miner on sparse datasets. FHM didn't terminate for more than 24 hours on Kosarak dataset for threshold value less than 0.7 $\%$ and mHUIMiner didn't terminate on Kosarak dataset for threshold value less than 0.75 $\%$. }
		\label{fig:real_sparse_perf_ufh}
	\end{center}
\end{figure}

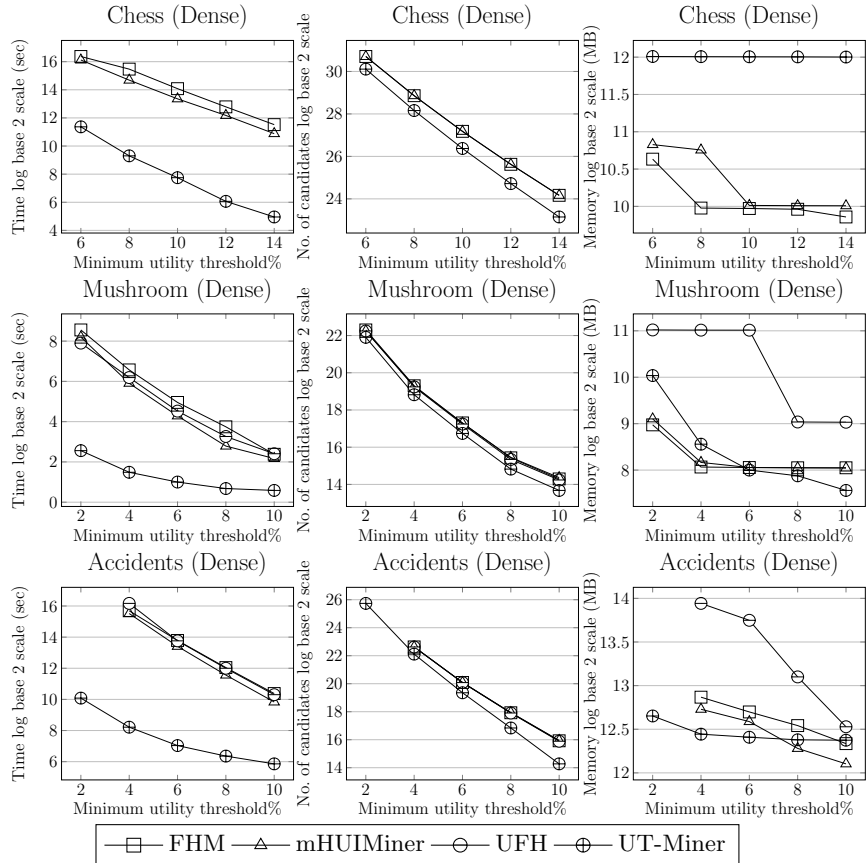
\begin{figure}
	\begin{center}
		\begin{tabular}{ccc}
			\centering
			
			\begin{minipage}{.27\linewidth}
				\begin{tikzpicture}[scale=0.45]
				\begin{axis}[
				title={\huge Chess (Dense)},
				xlabel={Minimum utility threshold$\%$ },
				ylabel={Time log base 2 scale (sec)},
				enlarge y limits=true,
					legend to name=t3,
					legend image post style={scale=1.3},
				legend style={at={($(0,0)+(1cm,1cm)$)},
					legend columns=4,fill=none,draw=black,anchor=center,align=center},
				ymajorgrids=true,
				mark size=5pt,
				label style={font=\Large},
                tick label style={font=\Large} 
				]
				
				\addplot[
				color=black,
				mark=square,
				style=solid,
				]
				coordinates {
					(6,{log2(84600)})(8,{log2(45391)})(10,{log2(17325)})(12,{log2(7066)})(14,{log2(2943)})
				};
				\addlegendentry{FHM}
				\addplot[
				color=black,
				mark=triangle,
				style=solid,
				]
				coordinates {
					(6,{log2(71136)})(8,{log2(26381)})(10,{log2(10486)})(12,{log2(4630)})(14,{log2(1888)})
				};    
				\addlegendentry{mHUIMiner}
				\addplot[
				color=black,
				mark=oplus,
				style=solid,
				]
				coordinates {
					(6,{log2(2638)})(8,{log2(631)})(10,{log2(215)})(12,{log2(67)})(14,{log2(31)})
				};    
				\addlegendentry{UT-Miner}
				
				\end{axis}
				
				\end{tikzpicture}
			\end{minipage}
			\hspace{0.3cm}
			
			&
			
			\begin{minipage}{.27\linewidth}
				\begin{tikzpicture}[scale=0.45]
				\begin{axis}[
				title={\huge Chess (Dense)},
				xlabel={Minimum utility threshold$\%$ },
				ylabel={No. of candidates log base 2 scale},
				enlarge y limits=true,
					legend to name=t3,
					legend image post style={scale=1.3},
				legend style={at={($(0,0)+(1cm,1cm)$)},
					legend columns=4,fill=none,draw=black,anchor=center,align=center},
				ymajorgrids=true,
				mark size=5pt,
				label style={font=\Large},
                tick label style={font=\Large} 
				]
				
				\addplot[
				color=black,
				mark=square,
				style=solid,
				]
				coordinates {
					(6,{log2(1750495580)})(8,{log2(487058997)})(10,{log2(152001239)})(12,{log2(51811065)})(14,{log2(18831146)})
				};
				\addlegendentry{FHM}
				\addplot[
				color=black,
				mark=triangle,
				style=solid,
				]
				coordinates {
					(6,{log2(1750490291)})(8,{log2(487058058)})(10,{log2(152001185)})(12,{log2(51811142)})(14,{log2(18831366)})
				};    
				\addlegendentry{mHUIMiner}
				\addplot[
				color=black,
				mark=oplus,
				style=solid,
				]
				coordinates {
					(6,{log2(1159148625)})(8,{log2(299788398)})(10,{log2(86891895)})(12,{log2(27602414)})(14,{log2(9289803)})
				};    
				\addlegendentry{UT-Miner}
				
				\end{axis}
				
				\end{tikzpicture}
			\end{minipage}
			\hspace{0.3cm}
			&
			\begin{minipage}{.27\linewidth}
				\begin{tikzpicture}[scale=0.45]
				\begin{axis}[
				title={\huge Chess (Dense)},
				xlabel={Minimum utility threshold$\%$ },
				ylabel={Memory log base 2 scale (MB)},
				enlarge y limits=true,
					legend to name=t3,
					legend image post style={scale=1.3},
				legend style={at={($(0,0)+(1cm,1cm)$)},
					legend columns=4,fill=none,draw=black,anchor=center,align=center},
				ymajorgrids=true,
				mark size=5pt,
				label style={font=\Large},
                tick label style={font=\Large} 
				]
				
				\addplot[
				color=black,
				mark=square,
				style=solid,
				]
				coordinates {
					(6,{log2(1588)})(8,{log2(1008)})(10,{log2(1004)})(12,{log2(997)})(14,{log2(928)})
				};
				\addlegendentry{FHM}
				\addplot[
				color=black,
				mark=triangle,
				style=solid,
				]
				coordinates {
					(6,{log2(1818)})(8,{log2(1729)})(10,{log2(1033)})(12,{log2(1031)})(14,{log2(1030)})
				};    
				\addlegendentry{mHUIMiner}
				\addplot[
				color=black,
				mark=oplus,
				style=solid,
				]
				coordinates {
					(6,{log2(4121)})(8,{log2(4115)})(10,{log2(4110)})(12,{log2(4107)})(14,{log2(4100)})
				};    
				\addlegendentry{UT-Miner}
				
				\end{axis}
				
				\end{tikzpicture}
			\end{minipage}
			\\
			\begin{minipage}{.27\linewidth}
				\begin{tikzpicture}[scale=0.45]
				\begin{axis}[
				title={\huge Mushroom (Dense)},
				xlabel={Minimum utility threshold$\%$ },
				ylabel={Time log base 2 scale (sec)},
				enlarge y limits=true,
					legend to name=t3,
					legend image post style={scale=1.3},
				legend style={at={($(0,0)+(1cm,1cm)$)},
					legend columns=4,fill=none,draw=black,anchor=center,align=center},
				ymajorgrids=true,
				mark size=5pt,
				label style={font=\Large},
                tick label style={font=\Large} 
				]
				
				\addplot[
				color=black,
				mark=square,
				style=solid,
				]
				coordinates {
					(2,{log2(376)})(4,{log2(95)})(6,{log2(31)})(8,{log2(13.4)})(10,{log2(5.2)})
				};
				\addlegendentry{FHM}
				\addplot[
				color=black,
				mark=triangle,
				style=solid,
				]
				coordinates {
					(2,{log2(294)})(4,{log2(60)})(6,{log2(19.5)})(8,{log2(6.9)})(10,{log2(4.5)})
				};    
				\addlegendentry{mHUIMiner}
				\addplot[
				color=black,
				mark=halfcircle,
				style=solid,
				]
				coordinates {
					(2,{log2(239)})(4,{log2(73)})(6,{log2(23)})(8,{log2(9.6)})(10,{log2(5.3)})
				};    
				\addlegendentry{UFH}
				\addplot[
				color=black,
				mark=oplus,
				style=solid,
				]
				coordinates {
					(2,{log2(5.9)})(4,{log2(2.8)})(6,{log2(2)})(8,{log2(1.6)})(10,{log2(1.5)})
				};    
				\addlegendentry{UT-Miner}
				
				\end{axis}
				
				\end{tikzpicture}
			\end{minipage}
			\hspace{0.3cm}
			&
			\begin{minipage}{.27\linewidth}
				\begin{tikzpicture}[scale=0.45]
				\begin{axis}[
				title={\huge Mushroom (Dense)},
				xlabel={Minimum utility threshold$\%$ },
				ylabel={No. of candidates log base 2 scale},
				enlarge y limits=true,
					legend to name=t3,
					legend image post style={scale=1.3},
				legend style={at={($(0,0)+(1cm,1cm)$)},
					legend columns=4,fill=none,draw=black,anchor=center,align=center},
				ymajorgrids=true,
				mark size=5pt,
				label style={font=\Large},
                tick label style={font=\Large} 
				]
				
				\addplot[
				color=black,
				mark=square,
				style=solid,
				]
				coordinates {
					(2,{log2(5166651)})(4,{log2(645966)})(6,{log2(161844)})(8,{log2(44081)})(10,{log2(20093)})
				};
				\addlegendentry{FHM}
				\addplot[
				color=black,
				mark=triangle,
				style=solid,
				]
				coordinates {
					(2,{log2(4976967)})(4,{log2(633388)})(6,{log2(160608)})(8,{log2(44880)})(10,{log2(21116)})
				};    
				\addlegendentry{mHUIMiner}
				\addplot[
				color=black,
				mark=halfcircle,
				style=solid,
				]
				coordinates {
					(2,{log2(4975427)})(4,{log2(619386)})(6,{log2(155146)})(8,{log2(41551)})(10,{log2(19332)})
				};    
				\addlegendentry{UFH}
				\addplot[
				color=black,
				mark=oplus,
				style=solid,
				]
				coordinates {
					(2,{log2(3920969)})(4,{log2(462392)})(6,{log2(109436)})(8,{log2(28981)})(10,{log2(13102)})
				};    
				\addlegendentry{UT-Miner}
				
				\end{axis}
				
				\end{tikzpicture}
			\end{minipage}
			\hspace{0.3cm}
			&
			\begin{minipage}{.27\linewidth}
				\begin{tikzpicture}[scale=0.45]
				\begin{axis}[
				title={\huge Mushroom (Dense)},
				xlabel={Minimum utility threshold$\%$ },
				ylabel={Memory log base 2 scale (MB)},
				enlarge y limits=true,
					legend to name=t3,
					legend image post style={scale=1.3},
				legend style={at={($(0,0)+(1cm,1cm)$)},
					legend columns=4,fill=none,draw=black,anchor=center,align=center},
				ymajorgrids=true,
				mark size=5pt,
				label style={font=\Large},
                tick label style={font=\Large} 
				]
				
				\addplot[
				color=black,
				mark=square,
				style=solid,
				]
				coordinates {
					(2,{log2(503)})(4,{log2(268)})(6,{log2(266)})(8,{log2(265)})(10,{log2(265)})
				};
				\addlegendentry{FHM}
				\addplot[
				color=black,
				mark=triangle,
				style=solid,
				]
				coordinates {
					(2,{log2(550)})(4,{log2(287)})(6,{log2(266)})(8,{log2(265)})(10,{log2(264)})
				};    
				\addlegendentry{mHUIMiner}
				\addplot[
				color=black,
				mark=halfcircle,
				style=solid,
				]
				coordinates {
					(2,{log2(2075)})(4,{log2(2069)})(6,{log2(2068)})(8,{log2(525)})(10,{log2(523)})
				};    
				\addlegendentry{UFH}
				\addplot[
				color=black,
				mark=oplus,
				style=solid,
				]
				coordinates {
					(2,{log2(1050)})(4,{log2(377)})(6,{log2(256)})(8,{log2(235)})(10,{log2(189)})
				};    
				\addlegendentry{UT-Miner}
				
				\end{axis}
				
				\end{tikzpicture}
			\end{minipage}
			\\ 
			\begin{minipage}{.27\linewidth}
				\begin{tikzpicture}[scale=0.45]
				\begin{axis}[
				title={\huge Accidents (Dense)},
				xlabel={Minimum utility threshold$\%$ },
				ylabel={Time log base 2 scale (sec)},
				enlarge y limits=true,
					legend to name=t3,
					legend image post style={scale=1.3},
				legend style={at={($(0,0)+(1cm,1cm)$)},
					legend columns=4,fill=none,draw=black,anchor=center,align=center},
				ymajorgrids=true,
				mark size=5pt,
				label style={font=\Large},
                tick label style={font=\Large} 
				]
				
				\addplot[
				color=black,
				mark=square,
				style=solid,
				]
				coordinates {
					(4,{log2(52995)})(6,{log2(14123)})(8,{log2(4222)})(10,{log2(1329)})
				};
				\addlegendentry{FHM}
				\addplot[
				color=black,
				mark=triangle,
				style=solid,
				]
				coordinates {
					(4,{log2(46866)})(6,{log2(10874)})(8,{log2(3025)})(10,{log2(912)})
				};    
				\addlegendentry{mHUIMiner}
				\addplot[
				color=black,
				mark=halfcircle,
				style=solid,
				]
				coordinates {
					(4,{log2(73605)})(6,{log2(13878)})(8,{log2(4093)})(10,{log2(1263)})
				};    
				\addlegendentry{UFH}
				\addplot[
				color=black,
				mark=oplus,
				style=solid,
				]
				coordinates {
					(2,{log2(1084)})(4,{log2(297)})(6,{log2(131)})(8,{log2(82)})(10,{log2(58)})
				};    
				\addlegendentry{UT-Miner}
				
				\end{axis}
				
				\end{tikzpicture}
			\end{minipage}
			\hspace{0.3cm}
			&
			\begin{minipage}{.27\linewidth}
				\begin{tikzpicture}[scale=0.45]
				\begin{axis}[
				title={\huge Accidents (Dense)},
				xlabel={Minimum utility threshold$\%$ },
				ylabel={No. of candidates log base 2 scale},
				enlarge y limits=true,
					legend to name=t3,
					legend image post style={scale=1.3},
				legend style={at={($(0,0)+(1cm,1cm)$)},
					legend columns=4,fill=none,draw=black,anchor=center,align=center},
				ymajorgrids=true,
				mark size=5pt,
				label style={font=\Large},
                tick label style={font=\Large} 
				]
				
				\addplot[
				color=black,
				mark=square,
				style=solid,
				]
				coordinates {
					(4,{log2(6485727)})(6,{log2(1115234)})(8,{log2(249023)})(10,{log2(62149)})
				};
				\addlegendentry{FHM}
				\addplot[
				color=black,
				mark=triangle,
				style=solid,
				]
				coordinates {
					(4,{log2(6489046)})(6,{log2(1117992)})(8,{log2(251142)})(10,{log2(63282)})
				};    
				\addlegendentry{mHUIMiner}
				\addplot[
				color=black,
				mark=halfcircle,
				style=solid,
				]
				coordinates {
					(4,{log2(6468348)})(6,{log2(1108575)})(8,{log2(245828)})(10,{log2(60794)})
				};    
				\addlegendentry{UFH}
				\addplot[
				color=black,
				mark=oplus,
				style=solid,
				]
				coordinates {
					(2,{log2(55934920)})(4,{log2(4539821)})(6,{log2(669961)})(8,{log2(117155)})(10,{log2(19825)})
				};    
				\addlegendentry{UT-Miner}
				
				\end{axis}
				
				\end{tikzpicture}
			\end{minipage}
			\hspace{0.3cm}
			&
			\begin{minipage}{.27\linewidth}
				\begin{tikzpicture}[scale=0.45]
				\begin{axis}[
				title={\huge Accidents (Dense)},
				xlabel={Minimum utility threshold$\%$ },
				ylabel={Memory log base 2 scale (MB)},
				enlarge y limits=true,
					legend to name=t3,
					legend image post style={scale=1.3},
				legend style={at={($(0,0)+(1cm,1cm)$)},
					legend columns=4,fill=none,draw=black,anchor=center,align=center},
				ymajorgrids=true,
				mark size=5pt,
				label style={font=\Large},
                tick label style={font=\Large} 
				]
				
				\addplot[
				color=black,
				mark=square,
				style=solid,
				]
				coordinates {
					(4,{log2(7474)})(6,{log2(6647)})(8,{log2(5964)})(10,{log2(5168)})
				};
				\addlegendentry{FHM}
				\addplot[
				color=black,
				mark=triangle,
				style=solid,
				]
				coordinates {
					(4,{log2(6790)})(6,{log2(6156)})(8,{log2(4975)})(10,{log2(4400)})
				};    
				\addlegendentry{mHUIMiner}
				\addplot[
				color=black,
				mark=halfcircle,
				style=solid,
				]
				coordinates {
					(4,{log2(15744)})(6,{log2(13764)})(8,{log2(8780)})(10,{log2(5908)})
				};    
				\addlegendentry{UFH}
				\addplot[
				color=black,
				mark=oplus,
				style=solid,
				]
				coordinates {
					(2,{log2(6437)})(4,{log2(5571)})(6,{log2(5437)})(8,{log2(5326)})(10,{log2(5310)})
				};    
				\addlegendentry{UT-Miner}
				
				\end{axis}
				
				\end{tikzpicture}
			\end{minipage}
			\\
			
		\end{tabular}
		
		\ref{t3}
		\caption{Performance evaluation for FHM, mHUIMiner, UFH, and UT-Miner on dense datasets. FHM, mHUIMiner, and UFH didn't terminate for more than 24 hours on Connect dataset. FHM, mHUIMiner, and UFH didn't terminate for more than 24 hours on Accidents dataset at 2$\%$ threshold. The UFH algorithm ran out of memory on Chess dataset.   }
		\label{fig:real_dense_perf_ufh}
	\end{center}
\end{figure}

\begin{figure}
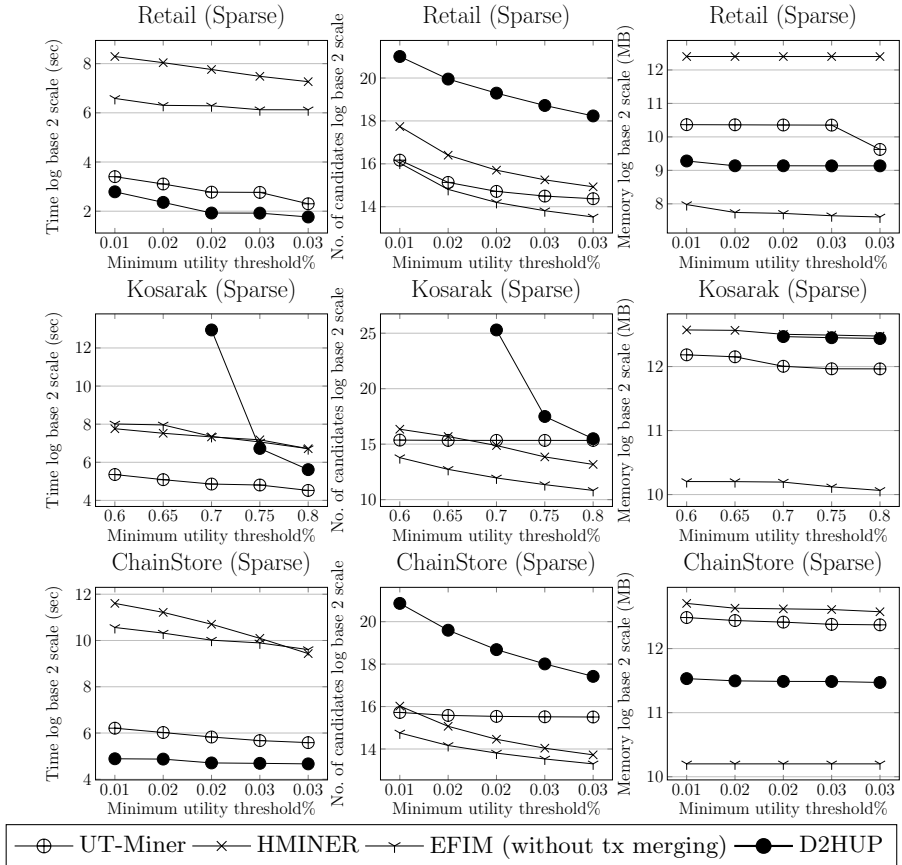

	\begin{center}

		
		\ref{t2}
		\caption{Performance evaluation for EFIM, D2HUP, HMiner and UT-Miner on sparse datasets. D2HUP didn't terminate for more than 24 hours on Kosarak dataset for threshold less than 0.7 $\%$. }
		\label{fig:real_sparse_perf_efim}
	\end{center}
\end{figure}

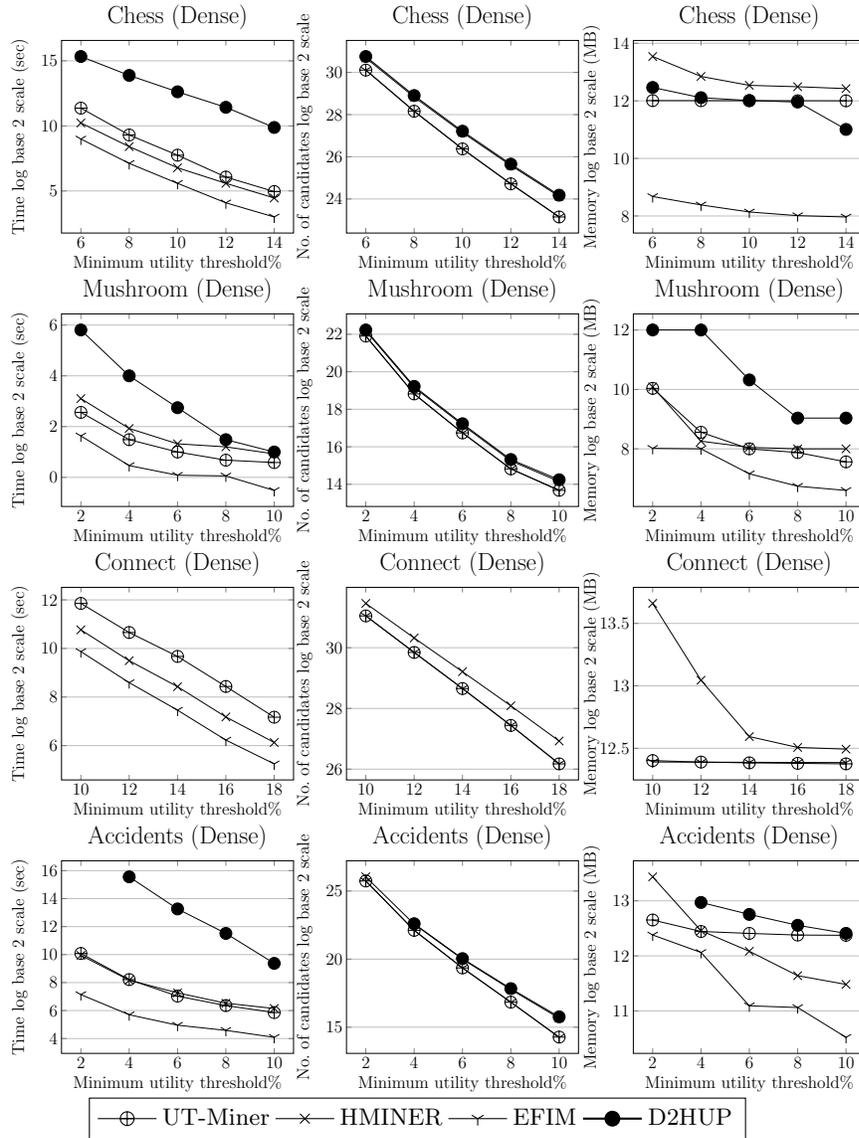
\begin{figure}
	\begin{center}
		\begin{tabular}{ccc}
			\centering
			
			\begin{minipage}{.27\linewidth}
				\begin{tikzpicture}[scale=0.45]
				\begin{axis}[
				title={\huge Chess (Dense)},
				xlabel={Minimum utility threshold$\%$ },
				ylabel={Time log base 2 scale (sec)},
				enlarge y limits=true,
					legend to name=t4,
					legend image post style={scale=1.3},
				legend style={at={($(0,0)+(1cm,1cm)$)},
					legend columns=4,fill=none,draw=black,anchor=center,align=center},
				ymajorgrids=true,
				mark size=5pt,
				label style={font=\Large},
                tick label style={font=\Large} 
				]
				
				\addplot[
				color=black,
				mark=oplus,
				style=solid,
				]
				coordinates {
					(6,{log2(2638)})(8,{log2(631)})(10,{log2(215)})(12,{log2(67)})(14,{log2(31)})
				};    
				\addlegendentry{UT-Miner}
				
				\addplot[
				color=black,
				mark=x,
				style=solid,
				]
				coordinates {
					(6,{log2(1196)})(8,{log2(339)})(10,{log2(110)})(12,{log2(48)})(14,{log2(22)})
				};
				\addlegendentry{HMINER}
				
				\addplot[
				color=black,
				mark=Mercedes star flipped,
				style=solid,
				]
				coordinates {
					(6,{log2(509)})(8,{log2(139)})(10,{log2(48)})(12,{log2(17)})(14,{log2(8)})
				};
				\addlegendentry{EFIM}
				
				\addplot[
				color=black,
				mark=*,
				style=thick,
				]
				coordinates {
					(6,{log2(41287)})(8,{log2(15081)})(10,{log2(6293)})(12,{log2(2752)})(14,{log2(942)})
				};
				\addlegendentry{D2HUP}
				\end{axis}
				
				\end{tikzpicture}
			\end{minipage}
			\hspace{0.3cm}
			
			&
			
			\begin{minipage}{.27\linewidth}
				\begin{tikzpicture}[scale=0.45]
				\begin{axis}[
				title={\huge Chess (Dense)},
				xlabel={Minimum utility threshold$\%$ },
				ylabel={No. of candidates log base 2 scale},
				enlarge y limits=true,
					legend to name=t4,
					legend image post style={scale=1.3},
				legend style={at={($(0,0)+(1cm,1cm)$)},
					legend columns=4,fill=none,draw=black,anchor=center,align=center},
				ymajorgrids=true,
				mark size=5pt,
				label style={font=\Large},
                tick label style={font=\Large} 
				]
				
				\addplot[
				color=black,
				mark=oplus,
				style=solid,
				]
				coordinates {
					(6,{log2(1159148625)})(8,{log2(299788398)})(10,{log2(86891895)})(12,{log2(27602414)})(14,{log2(9289803)})
				};    
				\addlegendentry{UT-Miner}
				
				\addplot[
				color=black,
				mark=x,
				style=solid,
				]
				coordinates {
					(6,{log2(1728424618)})(8,{log2(479473760)})(10,{log2(149065323)})(12,{log2(50584909)})(14,{log2(18294190)})
				};
				\addlegendentry{HMINER}
				
				\addplot[
				color=black,
				mark=Mercedes star flipped,
				style=solid,
				]
				coordinates {
					(6,{log2(1161454386)})(8,{log2(300484743)})(10,{log2(87133061)})(12,{log2(27692488)})(14,{log2(9325134)})
				};
				\addlegendentry{EFIM}
				
				\addplot[
				color=black,
				mark=*,
				style=thick,
				]
				coordinates {
					(6,{log2(1812818581)})(8,{log2(502257415)})(10,{log2(155683668)})(12,{log2(52772166)})(14,{log2(19009855)})
				};
				\addlegendentry{D2HUP}
				\end{axis}
				
				\end{tikzpicture}
			\end{minipage}
			\hspace{0.3cm}
			&
			\begin{minipage}{.27\linewidth}
				\begin{tikzpicture}[scale=0.45]
				\begin{axis}[
				title={\huge Chess (Dense)},
				xlabel={Minimum utility threshold$\%$ },
				ylabel={Memory log base 2 scale (MB)},
				enlarge y limits=true,
					legend to name=t4,
					legend image post style={scale=1.3},
				legend style={at={($(0,0)+(1cm,1cm)$)},
					legend columns=4,fill=none,draw=black,anchor=center,align=center},
				ymajorgrids=true,
				mark size=5pt,
				label style={font=\Large},
                tick label style={font=\Large} 
				]
				
				\addplot[
				color=black,
				mark=oplus,
				style=solid,
				]
				coordinates {
					(6,{log2(4121)})(8,{log2(4115)})(10,{log2(4110)})(12,{log2(4107)})(14,{log2(4100)})
				};    
				\addlegendentry{UT-Miner}
				
				\addplot[
				color=black,
				mark=x,
				style=solid,
				]
				coordinates {
					(6,{log2(11918)})(8,{log2(7377)})(10,{log2(5950)})(12,{log2(5755)})(14,{log2(5502)})
				};
				\addlegendentry{HMINER}
				
				\addplot[
				color=black,
				mark=Mercedes star flipped,
				style=solid,
				]
				coordinates {
					(6,{log2(409)})(8,{log2(334)})(10,{log2(282)})(12,{log2(258)})(14,{log2(250)})
				};
				\addlegendentry{EFIM}
				
				\addplot[
				color=black,
				mark=*,
				style=thick,
				]
				coordinates {
					(6,{log2(5654)})(8,{log2(4423)})(10,{log2(4128)})(12,{log2(3982)})(14,{log2(2058)})
				};
				\addlegendentry{D2HUP}
				\end{axis}
				
				\end{tikzpicture}
			\end{minipage}
			\\
\end{tabular}
		
		\ref{t4}
		\caption{Performance evaluation for EFIM, D2HUP, HMiner and UT-Miner on dense datasets. D2HUP didn't terminate for more than 24 hours on Connect dataset and Accidents dataset at 2$\%$ threshold.    }
		\label{fig:real_dense_perf_efim}
	\end{center}
\end{figure}

\noindent \textbf{Comparison with FHM, mHUIMiner, and UFH: } We compare the performance of our UT-Miner algorithm with FHM, mHUIMiner, and UFH. The results on Sparse datasets is shown in Figure \ref{fig:real_sparse_perf_ufh}. For the Kosarak dataset, it can be observed that UT-Miner performs the best by taking the least execution time compared to other algorithms. FHM and mHUIMiner don't terminate for more than 24 hours on the Kosarak dataset at lower threshold values. UFH takes two to three times more execution time than UT-Miner for the Kosarak dataset. UT-Miner also consumes the least memory due to its reuse of the tree structure and not creating local trees during every recursive call. For the Retail and ChainStore datasets, we observe that the running time of FHM, and mHUIMiner increases sharply for lower thresholds. UT-Miner takes slightly less time compared to UFH for Retail and ChainStore datasets. The results on dense datasets is shown in Figure \ref{fig:real_dense_perf_ufh}. UT-Miner performs two to six orders of magnitude better in running time compared to FHM and mHUIMiner on the Chess dataset. UFH ran out of memory on the Chess dataset. For the Mushroom dataset, UT-Miner performs one to four orders of magnitude faster than UFH, FHM, and mHUIMiner. UT-Miner performs better than FHM, mHUIMiner, and UFH on Accidents dataset too. FHM, mHUIMiner, and UFH didn't terminate for more than 24 hours on the Connect dataset. Hence, results are not reported for the Connect dataset. UT-Miner performs better than algorithms belonging to the category of list-based algorithms and algorithms that utilize both tree and list data structures on sparse and dense datasets. \\ \\ \\ \\     
\vspace{-1mm}
\noindent \textbf{Comparison with EFIM, D2HUP, and HMINER: } The result on sparse datasets is shown in Figure \ref{fig:real_sparse_perf_efim}. For the kosarak dataset, UT-Miner takes the least running time compared to EFIM, D2HUP, and HMiner. The running time of D2HUP increases sharply with the decrease in the minimum utility threshold. D2HUP didn't terminate for more than 24 hours at a threshold of less than 0.7 $\%$. The running time of UT-Miner is very close to D2HUP on Retail and ChainStore datasets. The running time of EFIM and HMINER increases sharply at low thresholds. EFIM with transaction merging disabled performs better on sparse datasets compared to EFIM with transaction merging enabled, as observed in this paper \cite{efim}. We disable transaction merging while running EFIM on sparse datasets. No significant change in the running time was observed by disabling transaction merging for the HMINER algorithm as it uses a hash-table to implement merging efficiently. We ran HMINER across all sparse and dense datasets with transaction merging enabled. The results for dense datasets is shown in Figure \ref{fig:real_dense_perf_efim}. EFIM performs the best by taking the least running time and memory across all dense datasets. D2HUP didn't terminate for more than 24 hours on the Connect dataset. UT-Miner performs slightly than HMiner and D2HUP for the Mushroom dataset. HMINER performs better than UT-Miner for the Connect and Chess datasets. Our proposed algorithm UT-Miner performs well across sparse and dense datasets.

\noindent \textbf{Influence of the number of transactions on execution time: } We experiment by varying the number of transactions for a fixed utility threshold on the ChainStore and Accidents dataset to study the impact of scalability on the performance of different algorithms. ChainStore and Accidents datasets are used for the experiment. The result is shown in Figure \ref{fig:real_scalability_perf}. The running time, the number of generated candidates, and memory consumption increase with the number of transactions for all algorithms. FHM, mHUIMiner, D2HUP, and UFH didn't terminate on Accidents dataset for more than 24 hours when more than 60 $\%$ of the transactions is input to the algorithm. HMiner and UT-Miner perform equally well on the Accidents dataset. 
   
\begin{figure}
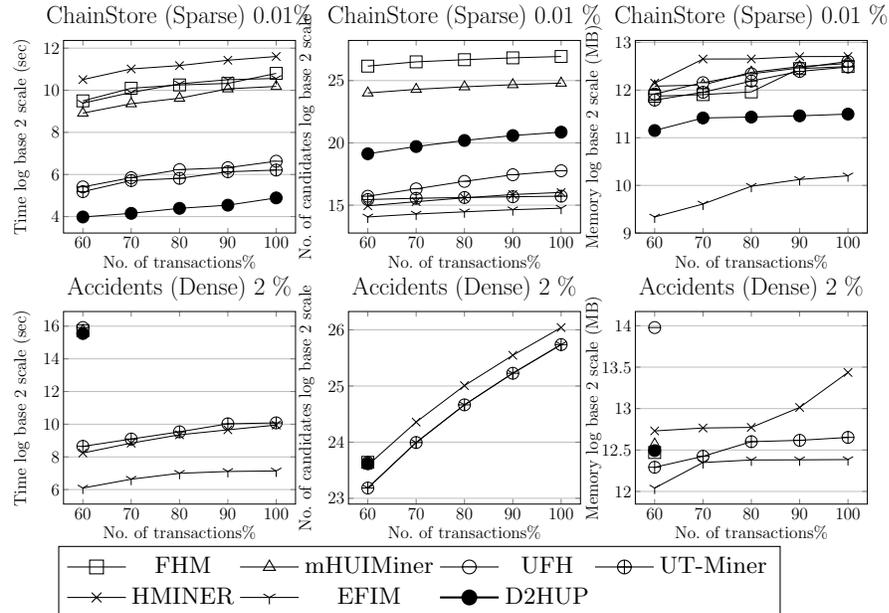

	\begin{center}
		%
		
		\ref{t2}
		\caption{Scalability experiment on ChainStore and Accidents dataset for 0.01$\%$ and 2$\%$ threshold respectively. FHM, mHUIMiner, D2HUP, and UFH didn't terminate for more than 24 hours on Accidents dataset when more than 60$\%$ of the transactions is input to the algorithms.  }
		\label{fig:real_scalability_perf}
	\end{center}
\end{figure}

\section{Conclusions}\label{sec:Conclusion and Future Work}
In this paper, we propose a novel data structure called Utility-Tree that stores the information about the transaction database compactly in the form of a HashMap structure with every node of the tree and a one-phase called UT-Miner with a lightweight projected database construction mechanism to mine high-utility itemsets from a transaction database. The tree structure stores the transaction database compactly and allows us to compute the $local\_list$ of the extensions for a given prefix efficiently using the HashMap data structure as the ancestors of the current prefix node are present in the transactions stored with the prefix node when the prefix is expanded in a bottom-up manner. Extensive experiments on several sparse and dense datasets against state-of-the-art algorithms reveal that UT-Miner is among the top-performing algorithms ranked according to ascending order of their execution time across sparse and dense datasets.

\bibliographystyle{unsrt}
\bibliography{mybibfile}
\end{document}